\documentclass[showpacs,preprintnumbers,amsmath,amssymb,nofootinbib]{revtex4}
\usepackage{graphicx}
\usepackage{epsfig}
\usepackage{bm}
\usepackage{amsfonts}
\newcommand{\Oo}{{\cal O}}
\newcommand{\Qo}{{\cal Q}}
\newcommand{\Po}{{\cal P}}
\newcommand{\Ro}{{\cal R}}
\newcommand{\Wo}{{\cal W}}
\newcommand{\Co}{{\cal C}}
\newcommand{\Eo}{{\cal E}}
\newcommand{\N}{{\cal N}}
\newcommand{\Jo}{{\cal J}}
\newcommand{\Wb}{{{\cal \hat W}}}
\newcommand{\erfc}{\text{erfc}}
\newcommand{\eq}{\begin{eqnarray}}
\newcommand{\eqx}{\end{eqnarray}}
\newcommand{\ba}{\begin{equation}}
\newcommand{\ea}{\end{equation}}
\newcommand{\f}[2]{\frac{#1}{#2}}

\newcommand{\n}{\nonumber \\}

\newcommand{\dl}{\delta}
\newcommand{\sig}{\sigma}
\newcommand{\iS}{\Sigma}

\newcommand{\Dl}{\Delta}
\newcommand{\cor}[1]{\left\langle{#1}\right\rangle}

\newcommand{\bit}{\begin{itemize}} 
\newcommand{\eit}{\end{itemize}} 
\newcommand{\ii}{\item} 
\newcommand{\fr}{\frac}

\def\tr{\tilde r}
\def\la{\label}
\def\nn{\nonumber \\}
\def\bi{\bibitem}
\def\g{\gamma}

\def\ka{\kappa}
\def\lam{\lambda}
\def\al{\alpha}
\def\als{\alpha_s}
\def\be{\beta}
\def\alp{\alpha'}

\def\vs{s}

\def\eps{\epsilon}
\begin{document}

\title{Dynamical entropy of dense QCD states}

\author{Robi Peschanski }
\email{robi.peschanski@cea.fr} \affiliation{Institut de Physique 
Th\'eorique,\\
CEA, IPhT, F-91191 Gif-sur-Yvette, France\\
CNRS, URA 2306 }
\today

\begin{abstract}
We discuss dense states of  QCD matter formed in high-energy hadronic and heavy-ion collisions from the point of view of statistical physics of non-equilibrium  processes. For this sake, we first  propose a formulation  of the {\it dynamical entropy} of dense QCD states in the ``saturation regime'' leading to a color glass condensate (CGC). The statistical physics description amounts to describe the modification of the color correlation length  with energy as a  $compression$ process for which nonequilibrium thermodynamic properties are applicable. We derive an expression  of the dynamical entropy in terms of  the rapidity evolution of the unintegrated gluon distributions in the colliding nuclei, verifying suitable positivity and irreversibility properties. We extend this approach to the initial pre-equilibrium (glasma) state of an heavy-ion collision. It  allows for a definition of the initial entropy before the evolution towards the hydrodynamic regime as a function of the glasma correlation length and an overlap parameter characterizing the  low-momentum spectrum of the glasma state. This initial entropy, by extension to the $\N=4$ SYM theory, is then matched as the key  input parameter to the strong coupling evaluation of thermalization towards the hydrodynamic regime based on the AdS/CFT correspondence. It thus allows to cast a bridge between the  weak and strong coupling phases of an heavy-ion reaction.
\end{abstract}

\pacs{12.38.-t,24.85.+p,25.75.Dw,05.70.-a} 
\maketitle

\section{Introduction}
\la{Introd}
It is commonly believed that statistical physics concepts may be relevant to describe the development and outcome of ultra-high energy collisions between incident hadrons or nuclei. For instance, the multi-production of particles  during such a high-energy reaction  has been related to   the entropy produced by the  collision. A statistical physics framework seems especially relevant for the central distribution of multiplicities in ultra-relativistic heavy-ion collisions, where  relativistic hydrodynamics, assumed during a transient evolution of the quark-gluon plasma (QGP), give a nice account  \cite{hydro} of the  observable features of the reaction.

However, the theoretical difficulty remains in relating these features to the basics of Quantum Chromodynamics (QCD), being the field-theoretical framework where, ultimately at least, those statistical physics features, and in particular the  entropy release, could be properly defined. Indeed, one faces both difficult  statistical-physics and field-theoretic problems, since the initial reaction medium  is  both far from equilibrium and submitted to a  time-dependent QCD evolution in a strong-coupling regime.

Recent interesting steps forward  made in this  direction are the following. In the context of strong gauge coupling studies using  the AdS/CFT correspondence \cite{adscft}, it was shown   \cite{przemek} that starting from various initial conditions, the thermalization leading to  a hydrodynamic behavior of a strongly-coupled gauge theory plasma essentially depends on an ``initial entropy'' factor defined  within the gauge/gravity duality framework using  involved notions of black hole (BH)  thermodynamics. The hydrodynamic regime is identified with the long-time behavior of a receding BH \cite{janik}. In this case, the entropy release during the evolution towards the hydrodynamic regime may be associated with the difference between the  final hydrodynamic entropy and  the abovementionned ``initial  entropy''. This encouraging result has yet some limitations. It applies to a maximally supersymmetric theory in the strong coupling regime, while one is in practice dealing with QCD with  weak-coupling initial states in the context of the Color Glass Condensate \cite{CGC} and the formation of an initial $glasma$ \cite{glasma}. Hence the problem is open to make the matching between a weak coupling initial phase and the strong coupling scenario of thermalization. Our goal is  to know how to define properly the ``initial  entropy'' for dense QCD states already at weak coupling.

In this context,
 a recent paper \cite{kutak} introduces the notion of a thermodynamical entropy associated with the production of gluons in the saturation regime of  dense  CGC initial states in proton-proton collisions\footnote{A different approach to entropy production  in heavy-ion collisions can be found in Refs.\cite{muller}.}. It is motivated by the relation between the saturation scale (denoted by $Q_s$ in transverse momentum) with a temperature of thermally produced hadrons and a dynamically generated mass, itself also related to the saturation scale $Q_s.$ This, together with the expected relation between entropy and the inclusive gluon production in a dense-dilute hadron-hadron collision, 
leads to an estimate of the entropy  of a dense  CGC initial state  in the framework of the Golec-Biernat W\"usthoff model \cite{GBW} for the QCD unintegrated gluon distribution (UGD).

Ref.\cite{kutak}  is a stimulating approach, since it proposes a $macroscopic$  approach to the entropy of a CGC ensemble of gluons, in a thermodynamic context. It may be  hindered by the assumptions of thermodynamic equilibrium, such as the existence of a temperature and  an entropy which are not yet well defined in principle in a perturbative QCD framework which seems to govern far-from-equilibrium processes where these notions are not standard. However, the notion of $dense$ QCD states, which is basic in the CGC approach, has specific features which may lead to a convenient statistical physics description of the high-energy collision processes without making assumptions on thermodynamic equilibrium.  In some sense it would be useful to have a $microscopic$ definition of entropy, analogous to the Boltzmann statistical mechanic approach,  based on the QCD dynamics of the CGC. We will propose  such a notion of a {\it dynamical entropy} for dense QCD states in the present paper.

In the next section \ref{id}, we will give the definition of a ``dynamical entropy'' for dense QCD states. Then in section
\ref{model}, we will apply it to models of the CGC medium, and check the matching with the macroscopic approach of Ref.\cite{kutak}. In the next section \ref{trans}, we will propose an  extension of the dynamical entropy approach to the  initial $glasma$ phase of heavy-ion collisions and analyze its matching with the thermalization scenario based on the AdS/CFT correspondence. A final section \ref{conclu} is devoted to conclusions with both a summary of results and outlook.

\section{Dynamical entropy of CGC states}
\la{id}

Dense QCD states may be present in various physical situations. In this section, we will consider the CGC medium. It  plays a role in  Deep Inelastic Scattering (DIS) in the ``small-$x$'' range where it represents the quantum states of the target as seen by the virtual photon when the Bjorken variable $x\sim Q^2/e^Y\ll 1,$ $Q^2$ being the photon virtuality and $Y$ is the total rapidity range of the final states. It also appears in the description of ``dense-dilute'' collisions such as proton-nucleus at very high energies. Later on, we will extend our analysis to the initial medium   in ultra-relativistic heavy-ion reactions obtained by the collision of  two  boosted  nuclei in their center-of-mass frame. 

These dense QCD states may be theoretically described  within  QCD at weak coupling, through the  nonlinear energy evolution  starting from  a dilute partonic state. The QCD evolution with total rapidity $Y$ increases the density of partons  and finally reaches the saturation regime  leading to a CGC state. It is characterized among other features \cite{CGC}, by a limiting transverse size $R_s(Y)\sim 1/Q_s(Y),$ where $Q_s(Y)$ is the rapidity-dependent saturation momentum. $R_s(Y)$ is in fact  the color correlation length at  given rapidity. 

As  energy increases,  the parton density becomes high enough such that individual parton branching is compensated by their recombination. This  leads to many interesting properties which can be evaluated  still in the (resumed at leading or next-leading logarithms) weak coupling regime of QCD. Among them  ``geometric scaling''  states \cite{geom} that the unintegrated gluon distribution ${\phi}(k;Y)$ of a CGC medium  essentially depends only on one variable implying the saturation scale. It verifies a scaling  property as a function of a single variable $u=k^2R_s^2(Y)$, within some controlled approximation \cite{QCDnonlinear,QCDnonlinearbis}, namely
\begin{equation}
 {\phi}(k,Y)\ d^2k \ \sim\  {\phi}(u=k^2R_s^2)\ R_s^2\ d^2k  \ .
 \la{phiscaling}
\end{equation}
The balance between parton branching and recombination at  saturation  and geometric scaling will be the main property allowing for a statistical mechanic approach, despite the probable absence of   thermal equilibrium.
Let us consider  this QCD compensation mechanism between branching and recombination from the statistical physics point-of-view in  \cite{jarzy,jarz2006,crooksy,sasa,crooksjarz,lua,relative,kirone} (choosing some useful references inside an abundant recent statistical physics literature, see $e.g.$ \cite{kirone}).

The CGC distribution of gluons can be obtained by the evolution of a partonic state (essentially gluons)  from a given  rapidity $Y_1$ to a dense medium  at higher rapidity $Y_2,$ characterized by a  spectrum in transverse momentum $k$  defined by the UGD $\phi(k,Y_2).$ Using the formulation of  classical\footnote{Though obtained through a quantum field theory framework, the system of partons generated by the QCD evolution has been proven to be described at leading logarithm approximation as a classical branching and recombination process.} statistical physics, the partons forming the CGC medium may be interpreted as a $stationary$ state, since, at each value of rapidity gluon branching and recombination compensate each other in the saturation regime. The evolution in rapidity $Y_1\!\to\! Y_2$ inside the dense regime  can be considered as an ({\it a priori}  nonequilibrium) process driven by a ``dynamical parameter'', being  the energy 
increase $\exp{Y_1/2}\!\to\! \exp {Y_2/2}$ of the partonic system. 

Indeed, in the statistical physics framework, a similar situation is met when a distribution of stationary states is defined through a probability distribution ${\cal P}^{\rm Stat}(z;\lam),$ where $z$ is a general notation for the phase space, see \cite{sasa,kirone} (and for a brief reminder, Appendix \ref{A2}). By changing the dynamical parameter $\lam_1\!\to\!\lam_2,$ the system evolves   in time leading to a final distribution which can be compared to a new stationary distribution ${\cal P}^{\rm Stat}(z;\lam_2).$ A key property of this nonequilibrium process is the Hatano-Sasa identity involving both ${\cal P}^{\rm Stat}(z;\lam_{1,2}),$ see Eq.\eqref{saha}.

Indeed, we are to show that the branching-recombining mechanism, responsible for the {\it geometric-scaling} property of the unintegrated gluon distribution $\phi(k,R_s) \sim \phi(kR_s(Y))$ gives rise to a transverse-momentum probability distribution verifying the analogous of the Hatano-Sasa identity \eqref{saha}.

We are led to the following identification (see Appendix \ref{A2}):
\ba
z\rightsquigarrow k, \quad\lam \rightsquigarrow Y,\quad {\cal P}^{\rm Stat}(z;\lam)\ dz \rightsquigarrow {\cal P}(k,Y)\ d^2k,
 \la{sahadefs}
 \ea
where, using \eqref{phiscaling}, the transverse momentum probability distribution  of gluons in the CGC is defined as 
\begin{equation}
{\cal P}(k,Y)\ d^2k = \frac {\ \phi(k,Y)\ d^2k}{\int\! \phi(k,Y)\ d^2k}\ \Rightarrow \ {\cal P}(u)\ du\ , 
 \la{relate}
\end{equation}
the last relation coming from the geometric scaling form \eqref{phiscaling}. This definition ensures the normalization condition
\begin{equation}
 \int {\cal P}(k;Y)\ d^2k =  \int {\cal P}(u)\ du \equiv 1 \ .
 \la{norm}
\end{equation}

 It is now straightforward  to write the identity 
\eq
 \cor{\exp{-\left[\int_{Y_1}^{Y_2} \frac {d}{dY}\left\{\log{\cal P}(k,Y)\right\}\ dY\right]}}_{Y_2}
&=& \int  {\cal P}(k,Y_2)\ d^2k \times\exp{-\left[\log\left\{\f{{\cal P}(k,Y_2)}{{\cal P}(k,Y_1)}\right\}\right]} \nn
&\equiv & 
\int {\cal P}(k,Y_1)\ d^2k 
\equiv 1\ ,
 \la{proof}
\eqx
where the left-hand side correlator $\cor{\cdots}_{Y_2}$ is defined by averaging over the probability distribution in the final state at $Y_2$ and making use of  the probability normalization relation \eqref{norm} at $Y_1.$


As we shall demonstrate further on, Eq. \eqref{proof}
 appears  as the QCD version  of the Hatano-Sasa identity \eqref{saha} for a CGC ensemble of states. It is a general property of QCD  saturation and the CGC \cite{CGC} that the gluons organize themselves in cells of typical saturation scale size. This is made quite explicitly, when introducing the geometric scaling formulation of \eqref{phiscaling}. Hence, the rapidity increase is responsible of  the shrinkage of the color correlation length which appears as a $compression$ process understood as the mean-field result of the denser gluonic medium. It gives rise to a modification of the probability distribution of gluon momentum which verifies the relation \eqref{proof}. This is explicit from \eqref{phiscaling}, where the compression  is manifest in the variation of the saturation scale $R_s(Y_1) \to R_s(Y_2)< R_s(Y_1).$

Thanks to this interpretation,
and following the proposed identification \eqref{sahadefs}, 
we introduce the  the notion of {\it dynamical entropy} as follows. We define  the {\it dynamical entropy} density of  a CGC parton medium at rapidity $Y_2,$ coming from the QCD evolution in rapidity $Y_1 \!\to \!Y_2$  inside a  transverse area of  order of the initial color correlation size $R_s^2(Y_1)$ by 
\ba
\Sigma^{Y_1\to Y_2} = \cor{\log{\frac{{\cal P}(k,Y_2)}{{\cal P}(k,Y_1)}}}_{Y_2}\equiv \int {\cal P}(k,Y_2)\  d^2k\ \log\left\{{\frac{{\cal P}(k,Y_2)}{{\cal P}(k,Y_1)}}\right\}\ .
\la{entropy}
\ea
We  note that
in the statistical physics literature formulas similar to  \eqref{entropy} appear also (see $e.g.$ \cite{relative}) as the  {\it relative entropy} of a probability distribution w.r.t. a reference one (here ${\cal P}(k,Y_2)$ compared to ${\cal P}(k,Y_1)$). It amounts to quantify the ``amount of disorder'' provoked by the rapidity evolution $Y_1 \!\to \!Y_2$ of the CGC medium. This notion is well adapted to the CGC medium of gluons, for which the competition between branching and recombination is expected to increase with the densification of gluons.\eject
Let us list  the role and properties of the dynamical entropy formula \eqref{entropy}.\\
\newline \i) $Positivity.$ By construction, the  identity \eqref{proof},
using the well-known \cite{jensen} Jensen inequality $e^{\cor{X}}\le \cor{e^X},$ for any probabilistic distribution  here identified to $X \equiv -\log{\frac{{\cal P}(k,Y_2)}{{\cal P}(k,Y_1)}},$  leads to 
the positivity condition
\ba
\Sigma^{Y_1\to Y_2} =  \cor{\log{\frac{{\cal P}(k,Y_2)}{{\cal P}(k,Y_1)}}}_{Y_2}\ge\ -\log \cor{\exp{\left\{-\log\left[{\frac{{\cal P}(k,Y_2)}{{\cal P}(k,Y_1)}}\right]\right\}}}_{Y_2}=0\ ,
\la{positivity}
\ea
for all $Y_2 \ge Y_1.$ Hence, the familiar positivity condition is obtained for any increase in total rapidity.\\
 \newline ii) {\it Geometric scaling case in terms of non-equilibrium thermodynamics.}
 Considering now the geometric scaling property \eqref{phiscaling}, the identity \eqref{proof} may also been discussed in thermodynamic terms, following  the discussion and properties of Jarzynski identity, see Appendix \ref{A1}  and \eqref{principle}. Indeed, using the geometric scaling version of \eqref{phiscaling}, the identity \eqref{proof} can be rewritten in geometric-scaling variables where for  short notation $R_{1,2}\!\equiv\! R_1(Y_{1,2}).$ one gets
\ba
  \cor{ \exp{-\left\{\log{\frac{{\cal P}(R^2_2k^2)}{{\cal P}(R^2_1k^2)}}\right\}}}_{R_2} =\exp{\left\{-\log{\frac{R^2_2}{R^2_1}}\right\}}\equiv \ \frac{Q^2_s(Y_2)}{Q^2_s(Y_1)}\ ,
 \la{jarzapplication}
\ea
where we reintroduced the saturation momentum scales $Q_S(Y_{1,2})=1/R_s^2(Y_{1,2}).$ 

The relation \eqref{jarzapplication} can be put in tight analogy with the  Jarzynski identity \cite{jarzy} of the thermodynamics of far-from-equilibrium processes (see Appendix \ref{A1}) which writes
\ba
\cor{e^{-\Wo/T}}_B = e^{-(F_C-F_A)/T} \ ,
\la{jarzintrod}
\ea
where the average is made over  different realizations of a process driving
 rapidly an equilibrium state $A$ at temperature $T$ to an out-of-equilibrium
 state $B,$ which  thermalizes towards a new equilibrium state  $C$
(keeping the driving dynamical parameter constant at its new value) at the same temperature.

 The Jarzinsky identity relates the stochastic distribution of thermodynamical
 works in the process $A\!\to\! B$ to the free energy balance $\Dl F$ between 
the two equilibrium states $A\!\to\!C.$ Interestingly, the amount of 
dissipative work $\Wo^{Diff}\equiv {\Wo}- \Dl F$ during the process  
$A\!\to\! B$ is then related to the entropy production $\Dl S = 
 (\cor{\Wo}- \Dl F)/T\ge 0\,,$ if the state $B$ is able to relax towards the temperature $T,$ keeping the driving parameter constant.
 
 Comparing now  \eqref{jarzapplication} to \eqref{jarzintrod}, one realizes
 that the expression $\log{{R^2_2}/{R^2_1}}$ corresponds to the logarithm of a ratio of the available phase-space for  dimension $R^2_2$ $vs.$ $R^2_1.$ It is  the free energy 
change of the particle of an ideal gas contained inside a 2-dimensional ``box'' when the size changes 
by  compression $R_1\!\to\!R_2<R_1.$ Hence, a thermodynamic  interpretation of
 the QCD relation   \eqref{proof} is that the modification of
 the total rapidity $Y_1\!\to\!Y_2$ induces a modification of the  CGC ensemble of states with reduced saturation size $R_2$
(or equivalently increased mean momentum $Q_2$), resulting in  an entropy eventually generated by further on relaxation. One is  led to the analogy 
\eq
\log{\frac{{\cal P}(R^2_2k^2)}{{\cal P}(R^2_1k^2)}}
&\rightsquigarrow & 
\f\Wo {T}  \nn 
\log{\frac{R^2_1}{R^2_2}}
&\rightsquigarrow & 
\f{\Dl F} {T}  \nn 
\log{\frac{{\cal P}(k,Y_2)}{{\cal P}(k,Y_1)}} 
&\rightsquigarrow& 
\f{\Wo -\Dl F}{T}\equiv
\f{\Wo^{Diff}}{T} \nn
\Sigma^{R_1\!\to\! R_2} = \cor{\log{\frac{{\cal P}(k,Y_1)}{{\cal P}(k,Y_2)}} }_{Y_2} &\rightsquigarrow& \Dl S\ ,
\la{analogies}
\eqx
where $\Dl S$ is, in the thermodynamic context, the {\it entropy production} (by gluonic degree of freedom) due to the compression $R_1\!\to\!R_2$ when the system relaxes to a state at the same temperature $T$ as the initial one but within the restricted domain size $R_2.$

The last line of \eqref{analogies} is suggestive of a special relation between the {\it dynamical entropy} of a CGC state $\iS$ and the entropy production
after relaxation $\Dl S.$ We shall confirm this analogy for Gaussian gluon distributions, by an explicit calculation in the next section.\\
\newline iii) {\it Dynamical entropy of a physical CGC state.}
The dynamical entropy formula \eqref{entropy} remains written for a unit gluon degree of freedom within the color correlation area and  for an arbitrary rapidity evolution $Y_1\!\to\!Y_2$ within the QCD saturation domain. In order to go to CGC states one deals with in high-energy collisions between hadrons and/or nuclei, one is due to extend  the formalism to those  physical cases. 

Using the thermodynamic analogy, the right-hand side of the identity \eqref
{jarzapplication} can be interpreted as coming from the standard free energy 
of one gluon d.o.f. inside of an ideal gas of gluons confined in a 
radial transverse box of saturation size $R_s,$ as expressed in the second raw of \eqref{analogies}. In order to determine the overall 
dynamical entropy that we may associate to a physical CGC initial state ($e.g.$ hadronic or nuclear one), we will assume that the probability distribution of reference ${\cal P}(k,Y_1)$ considers an  initial ``transverse cell''  at rapidity 
$Y_1$  where $R_1$ is a typical color correlation length for hadronic matter (proton or nucleus) at rest. In fact, this reference distribution will not play a role for large enough rapidity.

Then, we consider the 
boosted CGC medium when  ${Y_1\!\to\! Y_2}, {R_1\!\to\! R_2},$ where $Y_2$ is the rapidity at which one observes ($e.g.$ by deep-inelastic or dense-dilute hadronic scattering)  the corresponding CGC state. 
In  this 
framework, by addition of the individual contributions \eqref{entropy}, the total dynamical entropy density $dS_{T}/dy$ for a
boosted  target CGC state will be the sum over all degrees of freedom, 
namely  color multiplicity $N_c^2-1$, standard \cite{standard}
 gluon occupation number $\sim 1/4\pi N_c\als$ in longitudinal  coordinate 
space. For the transverse degrees of freedom, one has to take into account the  average number $R_T^2/R_2^2$ of ``transverse cells'' at initial rapidity $Y_1,$  where $R_T$ is the hadronic or nucleus target size, 
and finally the average number  $\mu$ of gluonic degrees of freedom inside a  cell which will  be determined later on using the Gaussian models for the UGD's. 
\eq
\f{dS}{dy} = \f {N_c^2-1}{4\pi N_c\als}\cdot  \f {R_T^2}{R_1^2}\cdot \mu \cdot \iS^{R_1\!\to\! R_2} \ .
 \la{relatecompressgausstot}
\eqx
Gluon correlation effects beyond  the  ideal gas approximation  may induce correction factors, which we do not discuss in the present study.

\section{Application to Gaussian CGC models}
\label{model}


We want now to apply our concept of  dynamical entropy $cf.$ \eqref{entropy} to simple CGC models in order to describe its main properties in a concrete way. Another motivation is to confront our definition of  dynamical entropy for a physical CGC state Eq.\eqref{relatecompressgausstot} to the one proposed in Ref.\cite{kutak}, which comes from the computation of the amount of gluon production, starting from a dense-dilute configuration of initial protons. 

Let us apply our approach to Gaussian models of  unintegrated gluon distributions, in which we introduce a parameter generalization of the GBW model \cite {GBW} in order to clarify the contributions to the entropy. The  GBW model indeed is  used by  \cite{kutak} for its definition of entropy which we wish to compare with our formulation.
We thus consider  unintegrated gluon distributions ${\phi}(k,Y)$ 
 \eqref{phiscaling} with a Gaussian tail, namely 
\eq
 {\cal P}(k,Y)\ d^2k \Rightarrow  {\cal P}(u=R_s^2k^2)\ du &=&
\psi(u)\ \ e^{-u}\ {du} \n
&\sim &  {\Gamma^{-1}(\ka)}\ {u^{\ka-1}} \ e^{-u}\ {du}\ , 
 \la{relatecompressgauss}
\eqx
where the non-Gaussian prefactor $\psi(u),$ describing the behavior in the low-$k$ range of  gluon momentum, is  more simply parametrized by a  power-like term $u^{\ka-1}.$ We shall call $\ka$ the ``overlap parameter'' for reasons becoming clear later on in this section. Note that $\ka=2,$ which is the GBW value, corresponds to a dipole distribution verifying the transparency property at zero dipole size \cite{GBW}.

Definition \eqref {relatecompressgauss} is indeed a slight generalization of   the unintegrated gluon distribution  coming from the Golec-Biernat Wusthoff  dipole model \cite{GBW,kutak} obtained for the value  $\ka=2.$ 


Starting by implementing \eqref{relatecompressgauss} in the dynamical entropy definition 
\eqref{entropy}, one gets
\eq
\Sigma^{Y_1\!\to\! Y_2}  &\equiv& \cor{\log{\frac{{\cal P}(R^2_2k^2)}{{\cal P}(R^2_1k^2)}}-
\log{\f {R_1^2}{R_2^2}}}_{Y_2}
= \cor{k^2}_{Y_2}(R_1^2-R_2^2) - \cor{\log{\f {R_1^2\cdot\psi(R_1^2k^2)}{R_2^2\cdot \psi(R_2^2k^2)}}}_{Y_2}\ \ge 0\ ,
 \la{relatecompressgaussw}
\eqx
and, more explicitly for the $\ka$-dependent family \eqref{relatecompressgauss}
\eq
\Sigma^{Y_1\!\to\! Y_2}  
&=& \int \left[k^2(R_1^2-R_2^2)-\ka\log{\f {R_1^2}{R_2^2}}\right]\ \Po(k^2R_2^2)\ R_2^2 dk^2 \nonumber\\
&=& 
\cor{k^2}_{Y_2}(R_1^2-R_2^2)-\ka\log{\left({R_1^2}/{R_2^2}\right)}=
\ka \left\{ \left({R_1^2}/{R_2^2}-1\right)-\log {\left( {R_1^2}/{R_2^2}\right)}\right\}\ \ge 0\ . 
 \la{relatecompressgausswka}
\eqx
One can easily check the strict final inequality \eqref{relatecompressgausswka} for all $Y_2 > Y_1$, $i.e.$ ${R_2}<{R_1}$ while it goes smoothly to zero (as $\ka/2\cdot\left({R_1^2}/{R_2^2}-1\right)^2$) when ${R_2}\!\to\!{R_1}.$ 


Let us list  relevant properties of formula  \eqref{relatecompressgausswka}.

\bit
\ii
{\it Dynamical entropy of the CGC state.}
Using the relation \eqref{relatecompressgausstot}, we can compute the total entropy density $dS^{Y_1\!\to\! Y_2}/\pi R^2_Tdy$ for a  CGC state of overall transverse size $R_T.$ Assuming  the framework of an initial  distribution of independent transverse ``cells'' with size $R_1$, it is given by
\eq
\f 1{\pi R^2_T}\f{dS^{Y_1\!\to\! Y_2}}{dy} =  \f {N_c^2-1}{4\pi^2 N_c\als}\ \f 1{R_1^2}\cdot \mu \cdot  \iS^{Y_1\!\to\! Y_2}
=\ka\mu\ \f {N_c^2-1}{4\pi^2 N_c\als}\  \left\{ {Q_2^2}-{Q_1^2}
\left(1+\log {\f{Q_2^2}{Q_1^2}}\right)\right\} \ ,
 \la{relatecompressgausstotion}
\eqx 
expressed in terms of the saturation momentum dependence $Q_s=1/R_s.$ Note that at large ${Q_2^2}/{Q_1^2}$,  formula \eqref{relatecompressgausstotion} shows that the 
target dynamical entropy by unit of transverse area and rapidity becomes independent of the initial condition at $Y_1$ and is directly proportional to $Q_s^2.$ The logarithmic corrections  provide a smooth transition to quasi reversibility $dS\! \to\! 0$ when ${Q_2^2}/{Q_1^2}\!\to\! 1$.

\ii
{\it ``Microscopic'' $vs.$ ``macroscopic'' entropy: determination of the $\mu$ parameter.}
Let us now derive the dynamical CGC entropy density precisely for the case considered in Ref.\cite{kutak}, and check the matching with the expression obtained in \cite{kutak} from gluon production  in the dense-dilute collision case. We tend to call it  ``macroscopic'' entropy since it relies on  a derivation from   thermodynamic concepts. In our case, that  we call ``microscopic'' we proceed from a counting of the configurations in transverse momentum phase-space parametrized by the probability ${\cal P}(k,Y),$ from
 \eqref{relatecompressgauss}.

For $p\!-\!p$ scattering, considered in \cite{kutak}, we use, for simplicity,  \eqref{relatecompressgausstotion} for the GBW gluon distribution with $\ka\!=\!2.$  We find for the dynamical (``microscopic'') entropy
\ba
 \f 1{A_T}\f{dS}{dy} =  \f {2\mu (N_c^2-1)}{\pi^2 \als N_c}{Q_2^2}\left(1 + \Oo\left[\f{Q_1^2}{Q_2^2}
\log { \f {Q_2^2}{Q_1^2}}\right]\right)\ ,
 \la{result}
\ea
with the proton mean transverse area $A_T\equiv\pi R_T^2/4.$

This result correctly matches the one\footnote{The relevant formula (25) in \cite{kutak} reads, in our notations
\ba
\f 1{A_T}\f {dS}{dy} =  \f {3(N_c^2-1)}{\pi \als N_c}{Q_s^2} + cst.\ ,
 \la{resultkutak}
\ea
} obtained in the derivation of ref. \cite{kutak}  which also gives the same behavior as the first term in formula \eqref{result}, independent of the initial condition. Hence the two derivations of the entropy of a CGC medium are consistent with each other despite their marked difference. Indeed,  one comes from the gluon production process in a dilute-dense collision and the other is based on the configurations of the CGC medium itself. 

Moreover, for an exact matching between the two formulas, one is led to identify
\ba
\mu \equiv \f{3\pi}2\ ,
\la{mu}
\ea 
which gives a non trivial estimate of the number of effectively independent gluonic degrees of freedom within a given color correlation length $R_s.$ Interestingly, it comes as a pure number, which may be characteristic of the average number of gluons inside a colorless cluster of a CGC medium generated by saturation. It would be worth checking that this remains valid for more elaborated UGD's.

An issue has been recently raised \cite{avsar1,avsar2} about 
$k_\bot$-factorization in single inclusive particle ($e.g.$ gluon) production in the small-$x$ regime of QCD and its formulation in terms of UGD's. Since the results on the ``macroscopic'' entropy of \cite{kutak} are obtained  from a $k_\bot$-factorization formula for two UGD's in $p\!-\!p$ scattering, some care is asked to be required for the comparison with our definition of the ``microscopic'' CGC entropy. In particular, the $\mu$ parameter depends on the correct normalization of the $k_\bot$-factorization formula, which choice is discussed (and heavily criticized in the current literature) in \cite{avsar1,avsar2}. We have checked that, provided starting with the appropriate definition of the dipole gluon distribution \cite{marquet} which is identical to the one of \cite{avsar1,avsar2}, we find full consistency with the normalization of the $k_\bot$-factorization formula defined in \cite{avsar1,avsar2}.  Hence, the different normalization choices made in \cite{kutak} leaves unchanged its formula, see \eqref{resultkutak}, and thus our result \eqref{result} for the  ``macroscopic'' entropy.  

\ii
{\it Strong irreversibility of the formation of a CGC state.} It is interesting to note that, using the same  method, one can compute from \eqref{entropy} the dynamical entropy $\Sigma^{R_2\!\to\! R_1}$ associated with the
 backward process , $i.e$ the expansion  $R_2\!\to\! R_1 > R_2.$ By exchanging $R_2\!\leftrightarrow\! R_1$ in \eqref{relatecompressgausswka} one finds 
\ba
\Sigma^{R_2\!\to\! R_1} \equiv \cor{\log{\frac{{\cal P}(k,Y_1)}{{\cal P}(k,Y_2)}} }_{Y_1}=\ka \left\{\log{\left({R_1^2}/{R_2^2}\right)}- \left(1-{R_2^2}/{R_1^2}\right)\right\}\ \ge 0\ .
\la{relateexpandw}
\ea
It is easy to check  also in that case the strict inequality for ${R_1}>{R_2},$ with a same smooth behavior towards the equality  as ${R_2}\!\to\! R_1.$ However, one finds a  logarithmic behavior $\log{\left({R_1^2}/{R_2^2}\right)}$ by contrast with the quadratic power-like behavior\eqref{relatecompressgausswka}. In fact, in that case the main contribution to the dissipative work is due to the free energy term in  \eqref{analogies}, while the ``work'' contribution is almost zero. The interpretation is that ($cf.$ \cite{lua}) only few  gluons within the stochastic distribution of transverse momenta feel the expansion of the color correlation length. Hence the ``gluon compression'' mechanism is in general much more dissipative-and thus irreversible- than ``gluon expansion''. This is even more remarkable, when assuming a ``cyclic'' process, namely a compression ${R_1}\!\to\! R_2$ followed by the reverse expansion ${R_1}\!\to\! R_2.$ The overall balance of entropy $\Dl \iS^{\circlearrowleft}(R_1,R_2)$, following Eqs.(\ref{relatecompressgausswka},\ref{relateexpandw}) writes
\ba
\Dl \iS^{\circlearrowleft}(R_1,R_2)= \ka\  
  \f{\left({R_1^2-R_2^2}\right)^2}{R_1^2 R_2^2} \sim \ka  \f{R_1^2}{R_2^2}\ ,
\la{bilan}
\ea
where the last approximation is for ${R_1 \gg R_2}$ large. Formula \eqref{bilan} makes explicit the irreversible feature of the color correlation size changing processes\footnote{Such ``cyclic'' process could be eventually describing the entropy production obtained through the formation and subsequent relaxation of a CGC state, when evaluated at weak QCD coupling. It would be interesting to relate it to the entropy production through thermalization.}. 

From this, one can induce that the QCD saturation mechanism due to the high density of gluons is generating a large entropy and thus highly dissipative in thermodynamic language.  Hence the dynamical entropy  \eqref{entropy} seems to capture interesting irreversible physical features of the formation of dense QCD states, and eventually, through the cyclic process, their relaxation towards equilibrium.

\ii
{\it Comparison with a 1-dimensional ideal gas model moved by a piston.} 
It is quite instructive to compare the results for the  QCD two-dimensional dynamical entropy \eqref{entropy} with the non-relativistic one-dimensional ideal gas model of Ref.\cite{lua}. 
One considers the rapid action $L_i \leftrightarrows L_f < L_i$ of a  piston on a  longitudinal box containing an ideal gas of independent particles, starting with temperature $T$. One can compute the work distribution due to the piston  either by  expansion \cite {lua} or compression \cite{notes}
The piston model  is a well-known far-from-equilibrium example where the Jarzynski identity \eqref{jarzintrod} can be explicitly  verified, and one finds for the average dissipative work $\cor{\Wo}- \Dl F$ (and thus the dynamical entropy)
\eq
\iS^{L_1 \rightarrow L_2} &\equiv & \f1{T}\left\{\cor{\Wo}- \Dl F^{L_1 \rightarrow L_2}\right\} \propto  {L_1}/{L_2}-1 - \log{\left( {L_1}/{L_2}\right)}\nonumber \\
 \iS^{L_1 \leftarrow L_2} &\equiv & \f1{T}\left\{\cor{\Wo}- \Dl F^{L_2 \rightarrow L_1}\right\} \sim  - \Dl F^{L_2 \rightarrow L_1} \propto \log{\left( {L_1}/{L_2}\right)}\ .
\la{disspiston}
\eqx
It is tempting  to interpret our results \eqref{relatecompressgausswka} and \eqref{relateexpandw} as the radial analogue of the two equations of \eqref{disspiston}. In some sense, the QCD field-theoretical saturation mechanism can  be thought of  as due to multiple shocks of individual gluons on the effective  ``walls'' due to the shrinkage of the color correlation length, as a mean-field effect on an individual gluon reproducing the many-body interactions of gluons at saturation. Rapid expansion is much less entropy producing, since it creates almost no work, leaving only, as a dissipative mechanism, the logarithmic free-energy cost $- \Dl F.$ As a side remark, it is interesting to figure out that saturated gluons within a small color correlation ``cell'' may behave as non relativistic particles, if they acquire an effective mass $M \sim Q_s$ as suggested in \cite{kutak}.
\ii
{\it Role of the ``overlap factor'' $\ka.$}
The factor $\ka,$ which parameterizes the low-$k$ range of the gluon transverse momentum distribution ($cf.$ \eqref{relatecompressgauss}), appears as an overall multiplicative factor in our resulting formulas for the dynamical entropy (\ref{relatecompressgausswka},\ref{relatecompressgausstotion},\ref{relateexpandw}) without modifying the quadratic dependence in $Q_s^2.$   It is linked to the structure of the gluon probability distributions ${\cal P}(k,Y_2)\ vs.\ {\cal P}(k,Y_1),$ $cf.$ \eqref{relatecompressgausswka}. Indeed, a smaller ($resp.$ larger) value of $\ka$ corresponds to wider ($resp.$ less)   overlap between the two distributions  and thus lower  ($resp.$ higher) amount of dynamical entropy. It appears to be much dependent on the low-$p_T$ range of the UGD's.

A remark here is  adequate concerning the Weiszs\"acker-Williams gluon distribution, which is also discussed in the framework of UGD's \cite{marquet}.
The  Weiszs\"acker-Williams gluon distribution has the physical interpretation as the number density of gluons inside the hadron/nucleus in light-cone gauge, but is not appearing in the  cross-section observables \cite{marquet}. The main difference with the one used in \eqref{relatecompressgausstotion} (and also in \cite{kutak}) is the infra-red logarithmic behavior  for $k^2 \!\to\!0,$  would correspond to an overlap factor $\ka \approx 1$ in \eqref{relatecompressgausswka}.  In our definition  \eqref{relatecompressgausstotion}  of the dynamical entropy, we choose the ``dipole gluon distribution'' and formula \eqref{relatecompressgausstotion} for the following reasons: it corresponds to a suppression of the color radiation at null transverse momentum, and a neat maximum at the saturation momentum $Q_s$, which is in a good analogy with the classical notion of a gas of particles in a ``box'' of size $R_s \!=\!1/Q_s,$ which makes consistent the  aspects of classical far-from equilibrium statistical physics on which relies our definition. This is related to the fact that QCD dipoles are colorless states which follow  $classical$ dynamics, at least at leading logs order of the perturbative expansion.
\eit

\section {Extension to  the dense initial state of heavy-ion reactions.}
\la{trans}
The pre-equilibrium state of matter reached immediately after the heavy-ion  collision, is only a transient one. In the most common scenario, it emanates from the interaction of the two  CGC states representing the wave-function of the incident colliding particles, forming ``at time zero'' what is thought to be  in the literature the $glasma$ phase \cite{glasma}. Then it evolves towards the hydrodynamic QGP phase,  during a strikingly  short-time period, as indicated by simulations, which is referred to as plasma $thermalization.$  This transition is still the less understood part of the QCD plasma formation in heavy-ion reactions. 
We shall investigate how the notion of a dynamical entropy may provide some new light on the questions raised by  the rapid thermalization found in heavy-ion collisions.

The two incident  nuclei considered in the center-of-mass frame are boosted and thus can be described as two independent CGC states.
Hence,  our formalism may in principle be  applied to the initial CGC states of both nuclei.  
However, one has to take into account the $glasma$ state formed  from these CGC states at the interaction time. Hence,  the 
application of our method defined in the previous section requires some extension. In particular, the significant matching with the ``macroscopic entropy'' of Ref.\cite{kutak}, see previous section, was made considering ``dense-dilute'' collision while  the $glasma$ state corresponds to a ``dense-dense'' configuration in the overall center-of-mass frame. In the following, we shall make a proposal how one can extend the notion of dynamical entropy for the glasma and its impact on the thermalization problem.

\vspace{.4cm}
$\bullet$ {\it  Dynamical entropy of the glasma.}
\vspace{.4cm}

Let us consider the  glasma phase of Ref. \cite{lappi2}, as an application example of our approach.
Extending the dynamical entropy concept to the glasma defined in  \cite{lappi2} appears nontrivial, since one has to consider a  classical Yang-Mills field calculation taking as initial value   the result of the JIMWLK evolution for each nucleus. Those calculations lead to a prediction for the  gluon distribution in the glasma $\phi_{glasma}(p_\bot,y)$ where $p_\bot$ is the transverse gluon momentum and $y,$ the rapidity evolution away from the initial condition at $y=0$ , taken in \cite{lappi2} as the McLerran-Venugopalan model \cite{McV}. What is numerically observed is that the resulting glasma spectrum satisfies geometric scaling using the gluon ($i.e.$ QCD adjoint) saturation momentum $Q_s(Y).$ We are thus led to apply a similar  recipe as in the CGC case, with the difference that new  nonlinear effects due to the classical Yang-Mills calculations are present, especially in the low $p_\bot$ range. One thus may proceed as follows.

Defining a probability distribution, similarly to \eqref{relate},\begin{equation}
{\cal P}_{glasma}(p_\bot,Y)\ d^2p_\bot \equiv \frac {\ \phi_{glasma}(p_\bot,Y)\ d^2k}{\int\! \phi_{glasma}(p_\bot,Y)\ d^2p_\bot}\Rightarrow {\cal P}_{glasma}(u=p_\bot^2/Q_s^2)\ du  , 
 \la{relateg}
\end{equation}
one is led to propose for  the glasma dynamical entropy density (in a transverse cell of order $R \sim 1/Q_s(Y)$), for the evolution $Y_1\!\to\!Y_2$ 
 \ba
\iS_{glasma}^{Y_1\!\to\!Y_2} \equiv  \cor{\log{\frac{{\cal P}_{glasma}(p_\bot,Y_2)}{{\cal P}_{glasma}(p_\bot,Y_1)}}}_{Y_2}\ge\ 0\ .
\la{entropyfonctionalg}
\ea

Assuming for simplicity a Gaussian from of the distribution, we consider a rapidity evolution of the glasma state in the range where geometric scaling with saturation momentum $Q_s(Y)$ is  valid.
Taking also into account in a phenomenological way the observed depletion \cite{lappi2} of gluon radiation in the small range of $p_\bot,$  one may write a formula similar to  \eqref{relatecompressgauss} being used for the CGC states in section \ref{model}, namely
\begin{equation}
{\cal P}^{glasma}\ d^2p_\bot  = \psi(p_\bot /Q_s)\ e^{-p_\bot^2/Q_s^2}\ d^2p_\bot \approx \f 1 {\Gamma(\ka_{gl})} \left(\f {p_\bot^2} {Q_s^2}\right)^{\ka_{gl}-1} e^{- {p_\bot^2}/{Q_s^2}}\ d^2p_\bot\ , 
 \la{relateint}
\end{equation}  
where the last approximation comes from considering the same input as in Ref.\cite{kutak}, $i.e.$ \eqref{relatecompressgauss} and the $\rm{small-}p_\bot$regularization prefactor $\psi(p_\bot/Q_s))$ being approximated using an
appropriate overlap factor $\ka_{gl}.$
%
Applying now our definition  \eqref{entropy} we get
\ba
\iS_{glasma}^{Y_1\!\to\! Y_2} \sim \ka_{gl}\  {Q_2^2}/{Q_1^2}\ , 
 \la{ions}
\ea 
where, as usual $Q_i\equiv Q_s(Y_i).$
It would be interesting to make a more precise evaluation of the glasma dynamical entropy and compare it with the sum of entropies of the initial CGC states for the same rapidity increase. This require to go beyond the Gaussian toy model we are using and we postpone this to further study.

\vspace{.4cm}

$\bullet$ {\it Entropy production and thermalization in heavy-ion collisions}

\vspace{.4cm}

The pre-equilibrium state of matter reached immediately after the heavy-ion  collision, is only a transient one. It evolves, during a quite short time as indicated by simulations, towards the hydrodynamic QGP phase which will  give rise finally to hadronization. This stage, driving the interacting medium from the CGC initial nucleus states through the glasma formation  to the hydrodynamic behavior of the QGP, called $thermalization$  is not well understood, due to lack of a $microscopic$ theory of thermalization based on  QCD. It may require a strong coupling treatment of collisions in QCD which is  not yet theoretically available. Some hints have been obtained \cite{przemek} from the AdS/CFT correspondence for the similar problem formulated in the case of the maximally symmetric ${\N}\! =\! 4$ gauge field theory.

In this context, let us 
examine  what can be obtained from  the formulation of the dynamical entropy for the initial state before thermalization, using the previous results. For this sake, we will confront both $macroscopic$ and $microscopic$ points of view. 

From a $macroscopic$  point-of-view inspired by  out-of-equilibrium statistical physics, there exists an intimate relation between  the dynamical entropy of an initial state as defined in our approach and the entropy production
during  thermalization. This has already been discussed in the case of the dilute-dense configuration of the initial states, see section \ref {model}. Or goal now is to extend it to the dense-dense case which is typical for heavy-ion collisions in the central rapidity region.
Second we will consider a  $microscopic$ approach, namely the strong coupling scenario of Ref.\cite{przemek}, where the notion of an  ``initial entropy'' has been  introduced using a dual gravitational version of the ${\N}\! =\! 4$ gauge field theory. We will conjecture that its determination for QCD may be based on the dynamical entropy of the glasma state.

Coming back for this sake to  the Jarzynski  and Hatano-Sasa identities. A key  feature  is their interpretation in terms of  the entropy generated by a fast non-equilibrium process in a finite time, and released to the environment by relaxation. In the case of the Jarzynski identity \eqref{jarzintrod}, see Appendix \ref{A2}, and assuming a thermodynamic coupling with an outside heat bath at  temperature $T,$ the exponent is related to the amount of average dissipative work $\cor{\Wo^{diff}}$, itself connected to the production of entropy $\Dl S$ by the relation
\eq
\Dl S \equiv \f1{T}\cor{\Wo^{diff}}\equiv \f1{T}(\cor{\Wo}- \Dl F)\ge 0\ .
\la{diff}
\eqx
In practice, the produced entropy  comes   from the relaxation of the out-of-equilibrium state by freezing the external force acting on the system, which finally evolves towards thermal equilibrium at the initial value of the temperature $T.$   

This process is, among others, well exemplified by the exact results obtained for one-dimensional  piston model \cite{lua}, see Fig.1. Starting from the thermal  state of an ideal  gas at temperature 
$T$ in a box of initial length  $L_i$, the action of the piston, with 
speed $u$  during a short time $\Dl \tau_1,$ shrinks the box size till a final size
$L_i\!\to\!L_f=L-|u|\Dl\tau_1$ which is then kept fixed, leaving the system relax towards the initial temperature\footnote{Another interesting exact model is the $adiabatic$ expansion and compression of a dilute interacting gas \cite{crooksjarz}}. Indeed, the action of the piston gives rise to the far-from-equilibrium 
state of Fig.1 middle, with the particles of gas being far from 
thermal equilibrium since most part of them acquires  momentum and kinetic 
energy by shocks on the piston \cite {lua}. Then its thermalization through 
the production of entropy  makes the system relaxing to the thermal state at temperature $T$ 
and length $L_f,$ depicted in Fig.1 right.
 \begin{figure}[t]
\begin{center}
\mbox{\epsfig{figure=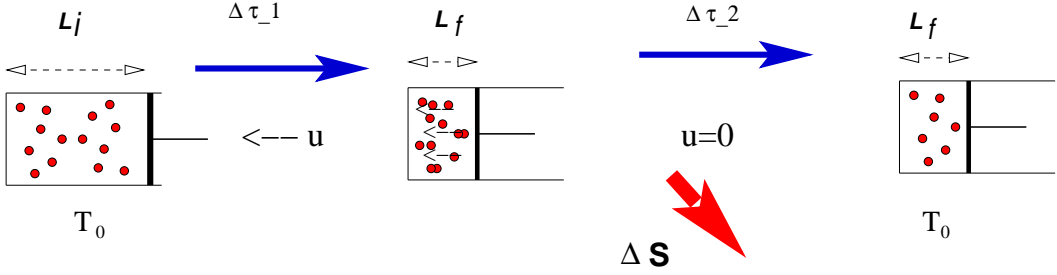,width=14cm,angle=0}}
\la{figtherm}
\caption{{\it Piston model: Compression and Thermalization.} Left: Thermal  state at temperature $T_0$ and length $L_i;$ Middle: Far-from equilibrium compressed state. Right: Thermalization through entropy production  to the heat bath and  relaxation to a denser thermal state at temperature $T$ and length $L_f.$}
\end{center}
\end{figure} 

It is interesting to note that the estimate \eqref{ions} of the dynamical entropy of the  initial glasma state appears as a simple extension of relation \eqref{diff} with 
\eq
\Dl S = \f1{T}\cor{\Wo^{diff}} \rightsquigarrow \Sigma^{Y_1\!\to\! Y_2}\ge 0\ ; \quad \f1{T}\cor{\Wo} \rightsquigarrow \ka_{gl}\ \left(\f{Q_{2}^2}{Q_1^2}-1\right)\ ; \quad \f1{T}\ \Dl F \rightsquigarrow \ka_{gl} \ \log {\left( \f{Q_{2}^2}{Q_1^2}\right)}\ ,
\la{diffqcd}
\eqx 
for  a two-dimensional version of the piston model. The physical
 interpretation is that the evolution of the glasma state is  obtained from the  $compression$ of  gluons confined within a box  of  color
 correlation length  $R_1\! \to \! R_2$ due to the increasing rate of branching and recombination of gluons at higher density. 

The precise determination of $\Dl S$
asks for a  mechanism incorporating  the expansion of the QCD matter, the thermalization process, the subsequent  hydrodynamic evolution and eventually   the hadronization of the QGP. While such a theoretical treatment is still unknown for QCD (beyond phenomenological approaches), a refined treatment of  thermalization through the hydrodynamic evolution could be obtained in  the strong coupling scheme of Ref.\cite{przemek} for the $\N\!=\!4$ SYM theory. Let us then consider this case in the context of our approach.
It has been shown that the boost-invariant proper-time expansion of the 
 $\N\!=\!4$ SYM plasma leads to a hydrodynamic stage  \cite{janik} with small 
viscosity over entropy ratio. Numerical solutions of the early-time evolution 
using the gravity dual allows one to examine the properties of the 
pre-equilibrium evolution, starting with various initial conditions which are 
not specified in  \cite{przemek}, and thus not related to the glasma states.
 We try and fill this gap in the sequel.

 The main common feature of the solutions of the strong coupling studies using the gauge/gravity duality, beyond the variety of initial  conditions, are the  short proper-time evolution towards the hydrodynamic stage, before which  they differ. Those regularities, such as entropy variation and the thermalization temperature
seem to only depend on the initial energy and entropy densities, namely
\eq
\epsilon(0)= N_c^2\cdot\f 38\pi^2\cdot T_{eff}^4(0)\quad ; 
\quad  \f{1}{A_T} \f{dS}{d\eta}=N_c^2\cdot\f 12\pi^2\cdot T_{eff}^2(0)\cdot s(0)
\la{densities}
\eqx
where one introduced  an ``initial effective temperature'' $T_{eff}(0)$ and an ``initial entropy'' density $s(0)$ by degree of freedom, both defined by \eqref{densities} from the corresponding energy and entropy densities by unit of space-time rapidity $\eta$ and transverse area $A_T.$ 

Note that the initial state {\it is not} at equilibrium and thus $T_{eff}(0)$ {\it is not} a physical temperature and  $s(0)$ {\it is not} the usual entropy at equilibrium. Following a notion of ``dynamical entropy'' in general relativity, it  is defined  $via$ the holographic correspondence as the element of the apparent horizon's area situated at geodesic distance at proper-time $\tau=0$ from the origin in the bulk space (see \cite{przemek} and references therein). 

Factorizing  the color degrees of freedom, we note that \eqref{densities} gives rise to the following relation
\eq
\f{1}{A_T}  \f{(3/4)\ {dS}/{d\eta}}{\ {\epsilon}(0)}=   s(0)\cdot T_{eff}^{-2}(0)\ ,
\la{densitiesratio}
\eqx
 valid for each degree of freedom of the initial state. Note that the correction factor $3/4$ is coming from the known ratio  (see $e.g.$ \cite{adscft}, section 3.6) between  the strong coupling entropy at finite temperature and the ideal gas one.  We shall not take into account this factor in the following, which deserves more study going beyond our context.
 
Considering now the corresponding densities for  dense QCD initial states, and for application, the Gaussian model \eqref{relateg}, we 
may write  formulas  in parallel with \eqref{densities} and  \eqref
{densitiesratio}.  For this sake, one starts from the formulas for the glasma 
energy density\footnote{For simplicity we use for the energy density in rapidity the standard
factor. For a more refined 
determination, see \cite{lappi}.} and for the dynamical entropy \eqref{ions}, 
with the standard counting of gluon degrees of freedom (assuming for sake of
 simplicity the same occupation number $1/(4\pi^2\als N_c)$ in rapidity as for
 CGC states, cancelled in the ratio). One obtains
\eq
\epsilon_{gl}= \f{N_c^2-1}{4\pi^2\als N_c}\cdot  Q_{gl}^4\quad ; \quad \f{1}{A_T} \f{dS_{gl}}{dy}=\ka_{gl}\cdot \mu_{gl}\cdot  \f{N_c^2-1}{4\pi^2\als N_c} \cdot Q_{gl}^2\ ,
\la{densitiesqcd}
\eqx
where $\eps_{gl}$ is the initial\footnote{There is a subtlety in the determination of $\eps_{gl}$ if using as an input the Weiszs\"acker-Williams definition of the UGD \cite{lappi} which diverges at $\tau=0.$ We here still use, for consistency of the Gaussian approach, the ``dipole'' UGD.} energy density of the glasma, $1/Q_{gl}=R_{gl}$ is the size of the glasma color correlation length, $\ka_{gl}$  the overlap factor in the Gaussian case \eqref{relateint} and $\mu_{gl}$ is the number of gluonic degrees of freedom in the color correlation ``cell'' of the glasma ($cf.$ \eqref{mu} for its determination for the CGC). One finally note that \eqref{densitiesqcd} gives rise to the following normalization independent relation
\eq
\f{1}{A_T} \f{{dS_{gl}}/{dy}}{{\epsilon}_{gl}(0)}=  \ka_{gl}\cdot \mu_{gl} \cdot\ Q_{gl}^{-2}\ .
\la{densitiesQCDratio}
\eqx

Matching \eqref{densitiesQCDratio} with \eqref{densitiesratio}  can be performed, since the proportionality of $T_{eff}(0)/Q_{gl}$  is expected from the scale invariance of both the  $\N\!=\!4$ SYM theory (from conformal invariance) and the dense state distributions in QCD at saturation
(from geometric scaling). Indeed, one finds:
\eq
T_{eff}(0) = \left[\f{s(0)}{\ka_{gl} \mu_{gl}}\right]^{1/2}\ Q_{gl}\ ,
\la{temp}
\eqx
which reads as a definition of the $effective$ temperature $T_{eff}(0)$ of the $\tau=0$ initial glasma state,
proportional to the size of the glasma correlation scale $Q_{gl}$ ($i.e.$ its inverse color correlation length). If, as expressed in Ref. \cite{Kharzeev}, the ratio $T_{eff}(0)/ Q_{gl}$ happens to be a universal $1/2\pi$ ratio connected to the so-called Unruh relation, we obtain that the initial entropy factor $s(0)$ increases with the overlap factor $\ka_{gl}$ and the $\mu_{gl}$ gluon occupation number in a ``cell''. This seems quite reasonable.

Now, in order to get a more quantitative estimate for the initial entropy $s(0)$ factor, we take as an example the  CGC case \cite{kutak}, one uses $\ka_{gl}=2,$ with $T_{eff}(0)/Q_{gl}= 1/(2\pi),$ from the determination of the temperature \cite{Kharzeev} through the Unruh property. In the Bjorken boost-invariant regime (\cite{hydro}, first reference) one can identify the space-like and energy-momentum definitions of rapidity $\eta \equiv y.$ Finally we assume the same value $\mu_{gl}=3\pi/2$ as in \eqref{mu} for the number of gluons by color cell. We then find 
\eq
s(0) \rightsquigarrow   \ka_{gl}\cdot\mu_{gl}\cdot  (T_{eff}(0)/Q_{gl})^2\ \Rightarrow\ \f 3{4\pi}\ .
\la{identify}
\eqx
where the last number comes from  the Gaussian form of a CGC state, taken as an example of the dense QCD state. Note that this number is in the ballpark of the range $0 <s(0)<1/2$ obtained in \cite{przemek}. It corresponds to a quite sizable entropy density.

More generally, the key order parameter $s(0)$ of the thermalization analysis  in  $\N\!=\!4$ SYM theory is in direct relation with  the ``overlap parameter'' $\ka_{gl},$ characterizing the strength of the dynamical entropy of the glasma state (in $\N\!=\!4$ SYM). This is consistent with the entropy increasing with decreasing overlap of probability distributions.

Recalling the main results of the numerical analyses showing \cite{przemek} that 
the system evolves quite rapidly towards a hydrodynamic behavior, even before isotropization occurs, the rate of entropy production $s_{f}\!-\!s(0)$ and thermalization proper-time (indeed, better to tell ``hydrodynamization time'') $\tau_{th}$ in units of $T_{eff}^{-1}(0)$ are driven by the value of $s(0).$ In our corresponding QCD evaluation \eqref{identify} for the Gaussian model, it is the value of the overlap parameter $\ka$ which is relevant. Higher is the dynamic entropy parameter $\ka$, $i.e.$ smaller is the overlap between the gluon probability distributions at $Q_2\ vs.\ Q_1,$ where $R_1=1/Q_1$ is taken for convenience of order of the standard color correlation length at rest. Then, greater is the initial entropy $s(0)$ and the subsequent increase of entropy $s_{f}\!-\!s(0)$ during thermalization. Following \cite{przemek}, the thermalization time is consequently shorter.
Hence, in our framework, it means that the thermalization time depends on the ``overlap parameter'', and thus on the strength of the dynamical entropy of the initial dense saturation state.


\section{Conclusions}
\la{conclu}

Let us give our conclusions. We have proposed a definition of a dynamical entropy  for dense QCD states of matter at high energies.
This definition is based on statistical physics tools  which have been shown to be valid for large classes of non-equilibrium processes. Our approach is inspired by  the Jarzynski and Hatano-Sasa fluctuation theorems \cite{jarzy,sasa} and is  applied to the formation of  dense QCD matter through an  energy-dependent evolution scheme defined in the QCD field theoretical framework at weak coupling. It describes  the evolution of the color correlation length  $R_1\!\to\! R_2,$  corresponding to the rapidity evolution   $Y_1\!\to\! Y_2$ of a dense QCD medium of gluons which is the result of gluon branching and recombination.

The derivation of an identity \eqref{proof} provides a guide  for the definition of a dynamical entropy functional verifying positivity, similar to the  $2^{nd}$ principle of irreversible thermodynamics.  It acquires an interesting physical picture when the gluon distribution of the dense QCD medium reaches the {\it geometric scaling} regime. In that stage, the boost-invariant distribution of gluons
can be described  as  transverse $cells$ whose radial size is  constrained by the collective effect of the others to be the $saturation\ scale\ R_s=1/Q_s(Y).$ The dynamical entropy is expressed as an overlap functional between the gluon distributions at different  total rapidities $Y_1$ and $Y_2>Y_1$ and thus saturation sizes   $R_s(Y_1)$ and $R_s(Y_2)<R_s(Y_1)$.

\subsection*{Summary}
Summarizing concretely our results:

\vspace{.1cm}
$\bullet \ Dynamical\ Entropy\  of
 \ CGC\ states:$ 

It is possible to define an entropy functional (by unit gluonic degree of freedom)
\ba
\iS^{Y_1\!\to\! Y_2} =  \cor{\log{\frac{{\cal P}(k,Y_)}{{\cal P}(k,Y_1)}}}_{Y_2} \equiv 
\int d^2k\ \log{\left\{\frac{{\cal P}(k,Y_2)}{{\cal P}(k,Y_1)}\right\}}\ {\cal P}(k,Y_2)\quad 
\ge\ 0\ ,
\la{entropyconclu}
\ea
where ${\cal P}(k,Y)$ is the  probability distribution of gluon transverse momentum at rapidity $Y.$ It is defined using  the QCD dipole unintegrated gluon distribution after normalization to unity. $\iS^{Y_1\!\to\! Y_2}$ is the dynamical entropy density (by transverse space ``cell'') during a far-from-equilibrium  ``compression'' process during which  the correlation length $R_1$ shrinks to $R_2 < R_1.$  

In the geometric scaling set-up,  the  color correlation length  of transverse size $R_s$ appears explicitly as
\ba
\iS^{Y_1\!\to\! Y_2} =  \cor{\log{\frac{{\cal P}(R^2_2k^2)}{{\cal P}(R^2_1k^2)}}}_{Y_2}-\log{\frac{R^2_1}{R^2_2}}\rightsquigarrow \f 1T \left\{{\cor{\Wo}^{R_1\!\to\! R_2}}- {{\Dl F}^{R_1\!\to\! R_2}}\right\}\ .
\la{analogiesconclu}
\ea
Hence,  \eqref{entropyconclu} in the geometric scaling form \eqref{analogiesconclu} admits the interpretation of a genuine entropy in a  thermodynamical  framework, see Appendix \ref{A1}, since the  $dissipative\ work$  performed during this process corresponds to  the difference between the average  work $\cor{\log{{{\cal P}(R^2_2k^2)}/{{\cal P}(R^2_1k^2)}}}_{Y_2}$ and the variation of the free energy $\log{{R_1^2}/{R_2^2}}$ between the initial and final  ``cells'', during the ``compression'' process ${R_1\!\to\! R_2}$ due to the increase of the gluon  density when the  total rapidity evolves ${Y_1\!\to\! Y_2}.$

In a non-thermodynamic framework, see Appendix \ref{A2}, where the CGC states are assimilated to non-equilibrium ``stationary states'', then $\iS^{R_1\!\to\! R_2}$ plays the role of a dynamical entropy acquired by the CGC ``initial stationary state'' from a rapidity  evolution and released through relaxation.
\vspace{.1cm}

$\bullet \ Application:\ the\  Gaussian\ model.$

As an example, we considered  dipole gluon densities of the form $\phi(k,R_1)\sim (kR_1)^{2(\ka-1)}\exp{\left(-k^2R_1^2\right)}$ generalizing the GBW 
model \cite{GBW}, 
for which one explicitly finds
\ba 
\iS^{R_1\!\to\! R_2}  =
\ka \left\{ \left({Q_2^2}/{Q_1^2}-1\right)-\log {\left( {Q_2^2}/{Q_1^2}\right)}\right\}\ \ge 0\ , 
 \la{gbwconclu}
\ea
where $Q_s({Y_1,Y_2})\equiv 1/R_{(1,2)},\ i.e.$ the saturation transverse momenta at the initial and final stage of the rapidity evolution. The parameter $\ka$ is an {\it overlap factor} characterizing the amount of overlap between the initial and final distributions, especially at low transverse momentum. Greater is $\ka,$ smaller is the overlap, and thus stronger is the dynamical entropy density generated by the evolution ${Y_1\!\to\! Y_2},$ due to the increasing number of branching and recombination QCD vertices involved in the process.

Interestingly, the result \eqref{gbwconclu} was proved to be identical for large $\Dl Y$, up to a factor $\mu \equiv 3\pi/2$, to the evaluation of the entropy variation of a CGC state coming from a $macroscopic$ formalism, $i.e.$ a thermodynamic interpretation \cite{kutak} of the distribution of gluons produced in a dense-dilute collision. The characteristic factor $\mu$ can   be interpreted as   the (average) number of gluonic degrees of freedom inside a transverse CGC ``cell''. To our knowledge, this is a new result coming from the comparison of two different but complementary approaches to the notion of entropy for a CGC medium. This appears to be  in conformity with the statistical physics interpretation, $via$ the comparison between $microscopic \ vs.\ macroscopic$ approaches,
 which relates the work fluctuations -here given the momentum distribution in the CGC state- to the entropy production - here, the gluon multiplicity produced in a ``dense-dilute'' collision probing the structure of the dense medium.
\vspace{.1cm}

\vspace{.1cm}
$\bullet$ {\it   Dynamical entropy  of the glasma 
in heavy-ion collisions.} 

We considered the class of models using the Color Glass Condensate formalism  describing the initial nuclei at the collision time and  leading to the $glasma$ as the initial interacting state of heavy-ion collisions.  Using the rapidity evolution of the glasma distribution of gluons, as $e.g.$ performed in Refs.\cite{lappi,lappi2}, it is possible, at least approximatively, to define the dynamical entropy for an rapidity variation $\dl y$ in a way similar to the CGC states. 

As an illustration, we used a phenomenological Gaussian model for  the probability distribution, namely
\begin{equation}
{\cal P}^{glasma}\ d^2p_\bot  =  \f 1 {\Gamma(\ka_{gl})} \left(\f {p_\bot^2} {Q_s^2}\right)^{\ka_{gl}-1} e^{- {p_\bot^2}/{Q_s^2}}\ d^2p_\bot\ , 
 \la{relateintconclu}
\end{equation}  
where, from the analysis of \cite{lappi2}, one  verifies geometric scaling with momentum scale $Q_s^2$ corresponding to the adjoint QCD representation, and a regularization at  smaller momentum due to the nonlinear effect of the classical Yang-Mills fields initiated by the initial CGC states. It is described here by a phenomenological value $\ka_{gl}$ of the overlap factor. One then gets in this example at leading order, for one gluonic degree of freedom,
\ba
\iS_{glasma}^{Y_1\!\to\! Y_2} \sim \ka_{gl}\ \left(\f  {Q_s(Y_2)}{Q_s(Y_1)}\right)^2\ . 
 \la{ionsconclu}
\ea 


$\bullet$ {\it   Hints for  thermalization:} 

The  evaluation through QCD calculations of the dynamical entropy for the initial state of heavy-ion collisions allows for interesting hints on the  problem of the transition towards the thermalized hydrodynamic stage of the Quark-Gluon Plasma. 

Noting that the expression \eqref{ionsconclu} for the dynamical entropy density of a glasma state can be a candidate for the ``initial entropy'' density $s(0)$ appearing as a key parameter in the strong coupling  thermalization mechanism \cite{przemek} of  the ADS/CFT correspondence, one finds the relation
\eq
s(0) \rightsquigarrow   \ka_{gl}\cdot \mu_{gl}\cdot \left\{ \f {T_{eff}(0)}{Q_{gl}}
\right\}^2\ ,
\la{identifyconclu}
\eqx
where $T_{eff}(0)/Q_{gl}$ is the ratio between the effective temperature (determined by the initial energy density) and the saturation scale of the glasma, expected to be a universal constant, independent of the total rapidity $Y.$ The parameter $\mu_{gl}$ is the number of gluonic degrees of freedom in a glasma ``cell''. Assuming for simplicity the same values for $\ka, \mu$ as for the Gaussian CGC distribution, we find values of $s(0)$ in the range compatible with the AdS/CFT
framework of Ref.\cite{przemek}.

It is interesting to see that \eqref{identifyconclu} gives a realization of the main physical idea behind the Jarzynski and Hatano-Sasa identities, namely the connection between the  work (and thus momentum) probability distribution
and the overall produced entropy by a far-from equilibrium process. Indeed, a large ($resp.$ small) overlap factor $\ka_{gl}$ means that the gluon momentum variations
induced by the  rapidity (and thus energy) evolution of a glasma state have been large  ($resp.$ small). Hence the contribution to the subsequent entropy (and thus particle) production is larger  ($resp.$ smaller). 

In practice, and if the input parameter values are taken from the CGC example in QCD and the AdS/CFT example from in $\N =4$ SYM theory, we find a rather large initial entropy density, in the range considered in  \cite{przemek}. It remains to see whether more realistic applications, both for QCD and AdS/CFT confirm this first approximation.

\subsection*{Outlook}

Many pending questions about non-equilibrium processes occurring during a high-energy heavy-ion collisions which may be discussed using the tools of non-equilibrium statistical mechanics, as we started to perform in the present study.
Let us mention some of the prospects in that direction.

\vspace{.1cm}
$\bullet \ Dynamical\ Entropy\  of\ dense\
 \ QCD\ states: theory$

Considering our application to QCD far-from-equilibrium processes, the  property of the stochastic identities ($e.g.$ \cite{jarzy,crooksy,sasa}  is the possibility of finding an expression for the entropy -usually related with the $final$ multiplicity- using the knowledge of the momentum distribution in the $initial$ state.
Hence, on the theoretical  level,  the question arises how the   properties of the dynamical entropy  are related to the basics of   the QCD evolution equations in the saturation setting, namely the BK or JIMWLK equations, see $e.g.$ \cite{CGC}. Indeed, they are  known  to lead to geometric scaling gluon distributions when the rapidity $Y$ increases, which is used in our derivation.  One more specific question is the role of the geometric scaling $violations,$ which are known to occur in the full solution of the QCD equations \cite{QCDnonlinear}. Hence, one may  ask what is the mathematical meaning of     the general expression of the dynamical entropy
\ba
\iS \equiv  \cor{\log{\frac{{\cal P}(k,Y_2)}{{\cal P}(k,Y_1)}}}_{Y_2}\ge\ 0\ ,
\la{entropyfonctional}
\ea
where ${\cal P}(k;Y_2)$ is proportional (with a suitable normalization) to the derivative of the gluon structure function $F(k,Y),$ solution of the BK/JIMWLK  QCD evolution equation as a function of the rapidity $Y$. This can be addressed as a question about integro-differential equations of F-KPP type which admit asymptotic traveling wave solutions, and  are the proper mathematical formulation of geometric scaling \cite{QCDnonlinear}.
\vspace{.1cm} 

$\bullet \ Dynamical\ Entropy\  of\ dense\
 \ QCD\ states: phenomenology$
 
On the phenomenological basis, it would be interesting  to make a more quantitative estimate of the various entropy contributions during all expected successive steps of a heavy-ion reaction using the phenomenological knowledge one has acquired about these states. Namely, we can usefully consider physical parameterizations of  CGC states and of the glasma phase, and of their evolution towards the hydrodynamic phase. These problems deserve  future studies.  

Finally, we may hope that the gap of knowledge yet existing for a proper general and operative formulation of  nonequilibrium processes in quantum field theories may be reduced thanks to studies on heavy-ion collisions. We conjecture that the existence of generalized { work identities}   may give some hints for elaborating such a  general framework, may be in relation  with the formalism of { Wilson lines and loops}, which is rooted in the bases of QCD, both at weak and strong coupling.
 
\vspace{.1cm} 

$\bullet\ The\ thermalization\ problem$ 
 
 The formalism and calculation presented in section \ref{trans} for the initial state entropy of an heavy-ion collision are in fact the {\it weak coupling} analogous of the strong coupling ones in the framework of the AdS/CFT correspondence performed in Ref.\cite{przemek}. A comparison between both approaches, see $e.g.$ \eqref{identifyconclu}, shows that a matching  can be obtained between the glasma initial state\footnote{We expect the glasma state to be similar for QCD and $\N=4$ SYM theories, which are both scale-invariant   in the leading-log perturbative domain.} and the subsequent evolution within the strong coupling scheme. the question now arises to eventually make a more direct comparison between the definition of entropy in those dynamical regimes: the one coming from the Shannon-type definition \eqref{entropyconclu} and the one coming from the dual gravitational theory related to the local area of the apparent horizon \cite{przemek}.  Our formalism may also be useful in the thermalization approach based on instabilities, $e.g.$ Ref.\cite{epelbaum}.

\appendix
\section{The Jarzynski identity}
\la{A1}
Generic identities have been derived when the evolution is fast and 
thus very far from equilibrium. These are the Jarzynski \cite{jarzy}
 and Hatano-Sasa \cite{sasa} (also Crooks \cite{crooksy}) fluctuation relations, which we now briefly describe in the text. 

It is well-known from statistical physics textbooks that expressions involving the average work $\cor{\Wo}$ 
performed by a statistical system verifies thermodynamic $inequalities$  
related to the second principle of thermodynamics, namely the increase of the 
total entropy for an isolated system. For instance, for the thermodynamic transition between 
two equilibrium states, one automatically verifies the celebrated second principle of thermodynamics
\ba
\Dl S =\f 1T (\cor{\Wo}- \Dl F)\ge 0\ \equiv \ \f 1T \times \cor{\Wo_{Diff}}\ge 0 \,,
\la{principle}
\ea
where $\Dl S$ is the increase of entropy, $T,$  the temperature, and $\Wo$ is
 the work performed in the medium $i.e.$ $\Dl E \equiv \cor{\Wo}$ the total 
energy change, $\Dl F$ the variation of the free energy of the system. In such
 a relation,  ${\Wo}- \Dl F={\Wo_{Diss}}$ is the 
{\it dissipative work} paid by the system during a nonequilibrium process. The
 equality to zero in \eqref{principle} is only for reversible processes, 
through slow dissipation-less evolution. However, as in particular shown in Ref. \cite{jarzy} there are necessarily fluctuations with 
${\Wo_{Diss}}< 0$ in the dissipative work distribution. 

To be a bit more precise in this short introduction, the Jarzinsky identity relates the stochastic distribution of thermodynamical
 works in the process $A\!\to\! B$ to the free energy balance $\Dl F$ between 
the two equilibrium states $A\!\to\!C.$ Interestingly, the amount of 
dissipative work $\Wo^{Diff}\equiv {\Wo}- \Dl F$ during the process  
$A\!\to\! B$ is then related to the entropy production $\Dl S = 1/{T}
 (\cor{\Wo}- \Dl F)\ge 0\,,$ if the state $B$ is able to relax towards the temperature $T,$ keeping the driving parameter constant.

The Jarzynski identity \cite{jarzy} reads
\ba
\cor{e^{-\Wo/T}} = e^{-\Dl F/T}\ . 
\la{jarzintrodA}
\ea
It is to realize that, while verifying the second principle $\cor{\Wo} -\Dl F = \Dl S\ge 0$ thanks to the Jensen identity \cite{jensen}, it implies   
deep properties on far-from-equilibrium fluctuations. Indeed,  it can be shown in many 
examples ($cf.$ \cite{jarz2006}) that a dominant contribution to \eqref{jarzintrodA} is 
given by  $rare$ out-of-equilibrium fluctuations very far from those dominating 
the average $\cor{\Wo}$ and thus $\Dl S$ in \eqref{principle}.

\section{The Hatano-Sasa identity}
\la{A2}
The nonequilibrium identities have a much larger range of applicability than the one yet considered initially \cite{jarzy}. Among the various extensions, we will select for our purposes the one \cite{sasa} which applies for purely non-equilibrium systems without reference to the notion of temperature. For this sake, one considers a situation where one has a system governed by a dynamical parameter $\lam$, such that at each value of $\lam,$  it corresponds to a $stationary$ phase space
spectrum with probability distribution ${\cal P}^{\rm Stat}(z;\lam) dz,$
where the variable $z$ describes the phase-space. Then the following equality holds \cite{sasa}
\begin{equation}
 \cor{\exp{-\left[\int_{\tau_1}^{\tau_2} d\tau \ \f {d\lam}{d\tau} \ \frac {\partial \log{\cal P}(z;\lam,\tau )}{\partial \lam}\right]}}_{\tau_2} \equiv \int dz\ \exp{-\left[\int_{\tau_1}^{\tau_2} d\tau \ \f {d\lam}{d\tau} \ \frac {\partial \log{\cal P}(z;\lam,\tau )}{\partial \lam}\right]}\times{\cal P}(z;\lam_2,\tau_2) \equiv 1\ ,
 \la{saha}
\end{equation}
where by definition ${\cal P}(z;\lam,\tau)\equiv {\cal P}^{\rm Stat}(z(\tau);\lam(\tau))$ is the stationary solution for the value $\lam(\tau)$ in  the ``frozen'' phase space variables at time $\tau.$ Note that the resulting identity is  $a\ priori$ not dependent on the arbitrary ``history'' $\f {d\lam}{d\tau}$ of the non equilibrium mechanism.
 
In fact, the quantity $-\log{\cal P}^{\rm Stat}(z;\lam)$ provides a candidate definition for a quantity playing the role, after integration in phase-space of a   ``dynamical entropy '' denoted $\iS$. Indeed, for the dynamical non-equilibrium process, where one starts with a state corresponding to $\lam_1$ to a state $\lam_2,$ the increase of `dynamical entropy'' gives rise to the following formula 
\ba
\Dl \iS^{\lam_1\!\to\!\lam_2}= \cor{\log{\f{{\cal P}^{\rm Stat}(z,\lam_1)}{{\cal P}^{\rm Stat}(z,\lam_2)}}}_{\lam_2}\equiv -\int \log\left\{\f{{\cal P}^{\rm Stat}(z,\lam_1)}{{\cal P}^{\rm Stat}(z,\lam_2)}\right\}\ \times\ {\cal P}^{\rm Stat}(z;\lam_2)\ dz\ .
\la{sahashannon}
\ea
In fact, our approach, see \eqref{proof}, Eq.\eqref{saha} is simply obtained by noting that there is no parameter dependence  on time, but only $via$ $\lam\rightsquigarrow Y$ which allows for an explicit integration in \eqref{sahashannon}. Moreover, geometric scaling ensures the identity \eqref{saha} since ${\cal P}^{\rm Stat}(z(\tau);\lam(\tau))$ depends only one one variable $u(\tau)\equiv z(\lam(\tau)).$

As can be noticed, a key feature is that no equivalent of a temperature  appears in the relation \eqref{saha}. In the Jarzynski case using Boltzmann factors, where temperature appears explicitly, the Hatano-Sasa identity is proven to reduce  to the original Jarzynski equation  \eqref{jarzintrodA}, namely
\ba
\cor{e^{-\f 1T (\Wo -\Dl F)}}\equiv \cor{e^{-\f 1T \Wo_{\rm diss}}} =1\,,
\la{jarzdiss}
\ea
where $\Wo^{\rm diss}$ is the dissipative part of the work.

\acknowledgements
We want to thank Kirone Mallick for useful explanations on the properties of Statistical Physics of far-from-equilibrium processes and in particular on the Jarzynski and Hatano-Sasa identities. Romuald Janik  and Krzysztof Kutak are warmly  thanked for works and  discussions which where much stimulating for our approach.


\begin{thebibliography}{0}

\bibitem{hydro}
  J.~D.~Bjorken,
   ``Highly Relativistic Nucleus-Nucleus Collisions: The Central Rapidity
  Region,''
  Phys.\ Rev.\  D {\bf 27}, 140 (1983).
  For reviews and references on hydrodynamics of the QGP, see
P.~F.~Kolb and U.~W.~Heinz,
  ``Hydrodynamic description of ultra-relativistic heavy-ion collisions,''
  arXiv:nucl-th/0305084.
  P.~Huovinen and P.~V.~Ruuskanen,
  ``Hydrodynamic Models for Heavy Ion Collisions,''
  Ann.\ Rev.\ Nucl.\ Part.\ Sci.\  {\bf 56}, 163 (2006)
  [arXiv:nucl-th/0605008].
  J.~Y.~Ollitrault,
  ``Relativistic hydrodynamics,''
  Eur.\ J.\ Phys.\  {\bf 29}, 275 (2008)
  [arXiv:0708.2433 [nucl-th]].

\bi{adscft} 
  For an introductory review and original references, see:
O.~Aharony, S.~S.~Gubser, J.~M.~Maldacena, H.~Ooguri and Y.~Oz,
  ``Large N field theories, string theory and gravity,''
  Phys.\ Rept.\  {\bf 323}, 183 (2000)
  [hep-th/9905111].

\bibitem{przemek}
  M.~P.~Heller, R.~A.~Janik and P.~Witaszczyk, ``The characteristics of thermalization of boost-invariant plasma from holography,''
  Phys.\ Rev.\ Lett.\  {\bf 108}, 201602 (2012)
  [arXiv:1103.3452 [hep-th]],\\
  ``A numerical relativity approach to the initial value problem in asymptotically Anti-de Sitter space-time for plasma thermalization - an ADM formulation,''
  Phys.\ Rev.\ D {\bf 85}, 126002 (2012)
  [arXiv:1203.0755 [hep-th]].

\bi{janik}
  R.~A.~Janik and R.~B.~Peschanski,
  ``Asymptotic perfect fluid dynamics as a consequence of AdS/CFT,''
  Phys.\ Rev.\  D {\bf 73}, 045013 (2006),
  ``Gauge / gravity duality and thermalization of a boost-invariant perfect
  fluid,''
  Phys.\ Rev.\  D {\bf 74}, 046007 (2006);\\
  See, $e.g.$ the review:
M.~P.~Heller, R.~A.~Janik and R.~Peschanski,
  ``Hydrodynamic Flow of the Quark-Gluon Plasma and Gauge/Gravity
  Correspondence,''
  Acta Phys.\ Polon.\  B {\bf 39}, 3183 (2008)
  [arXiv:0811.3113 [hep-th]].
  
\bibitem{CGC}
  F.~Gelis, E.~Iancu, J.~Jalilian-Marian, R.~Venugopalan,
  ``The Color Glass Condensate,''
  Ann.\ Rev.\ Nucl.\ Part.\ Sci.\  {\bf 60}, 463-489 (2010).
  [arXiv:1002.0333 [hep-ph]], for a recent review and original references.
  
  \bibitem{glasma}
 T.~Lappi and L.~McLerran,
  ``Some features of the glasma,''
  Nucl.\ Phys.\  A {\bf 772}, 200 (2006)
  [arXiv:hep-ph/0602189];\\
  L.~McLerran,
  ``The CGC and the Glasma: Two Lectures at the Yukawa Institute,''
  Prog.\ Theor.\ Phys.\ Suppl.\  {\bf 187}, 17-30 (2011).
  [arXiv:1011.3204 [hep-ph]], for a recent review and original references.

 \bi{kutak}
  K.~Kutak,
  ``Gluon saturation and entropy production in proton proton collisions,''
  Phys.\ Lett.\  B {\bf 705}, 217 (2011)
  [arXiv:1103.3654 [hep-ph]].

\bibitem{muller}
  B.~Muller and A.~Schafer,
  ``The decoherence time in high energy heavy ion collisions,''
  Phys.\ Rev.\  C {\bf 73}, 054905 (2006)
  [arXiv:hep-ph/0512100];\\ 
 R.~J.~Fries, B.~Muller and A.~Schafer,
  ``Decoherence and Entropy Production in Relativistic Nuclear Collisions,''
  Phys.\ Rev.\  C {\bf 79}, 034904 (2009)
  [arXiv:0807.1093 [nucl-th]];\\
  For a recent review:  B.~Muller and A.~Schafer,
  ``Entropy Creation in Relativistic Heavy Ion Collisions,''
  Int.\ J.\ Mod.\ Phys.\  E {\bf 20}, 2235 (2011)
  [arXiv:1110.2378 [hep-ph]].

\bi{GBW}
  K.~J.~Golec-Biernat and M.~Wusthoff,
  ``Saturation effects in deep inelastic scattering at low Q**2 and its
  implications on diffraction,''
  Phys.\ Rev.\  D {\bf 59}, 014017 (1998)
  [arXiv:hep-ph/9807513].


\bi{geom}
  A.~M.~Stasto, K.~J.~Golec-Biernat and J.~Kwiecinski,
  ``Geometric scaling for the total gamma* p cross-section in the low x
  region,''
  Phys.\ Rev.\ Lett.\  {\bf 86}, 596 (2001)
  [arXiv:hep-ph/0007192].

\bibitem{QCDnonlinear}
  S.~Munier and R.~B.~Peschanski,
  ``Geometric scaling as traveling waves,''
  Phys.\ Rev.\ Lett.\  {\bf 91}, 232001 (2003)
  [arXiv:hep-ph/0309177];
  ``Traveling wave fronts and the transition to saturation,''
  Phys.\ Rev.\  D {\bf 69}, 034008 (2004)
  [arXiv:hep-ph/0310357];
  ``Universality and tree structure of high energy QCD,''
  Phys.\ Rev.\  D {\bf 70}, 077503 (2004)
  [arXiv:hep-ph/0401215].
  
 \bibitem{QCDnonlinearbis}
 A.~H.~Mueller and D.~N.~Triantafyllopoulos,
  ``The energy dependence of the saturation momentum,''
  Nucl.\ Phys.\  B {\bf 640}, 331 (2002)
  [arXiv:hep-ph/0205167]. \newline
  D.~N.~Triantafyllopoulos,
   ``The energy dependence of the saturation momentum from RG improved BFKL
  evolution,''
  Nucl.\ Phys.\  B {\bf 648}, 293 (2003)
  [arXiv:hep-ph/0209121].

 
 \bibitem{jarzy}
C. Jarzynski,   ``A nonequilibrium equality for free energy differences''
Phys.\ Rev.\ Lett.\  {\bf 78}, 2690 (1997); ``Equilibrium free energy differences from nonequilibrium measurements: a master equation approach''; \\ For a review:
``Nonequilibrium work relations: foundations and applications'',
  Eur.\ Phys.\ J. \ B {\bf 64}, 331 (2008). 


\bi{jarz2006}
Phys.\ Rev.\  E {\bf 56}, 5018  (1997); 
``Rare events and the convergence of exponentially averaged work values''
Phys.\ Rev.\ {\bf E 73}, 046105 (2006), arXiv:cond-mat/0603185.


  \bi{crooksy}
G.E.Crooks, J.\ Stat.\ Phys.\ {\bf 90}, 1481 (1998);
  Phys.\ Rev.\  E {\bf 60},  2721 (1999).
  
 
\bi{sasa}
    T. Hatano, S. Sasa,    
    ``Steady State Thermodynamics of Langevin Systems'',  Phys.\ Rev.\ Lett.\  {\bf 86},  3463 (2001). 

\bi{crooksjarz}
    G.E. Crooks, C. Jarzynski,      
    ``On the work distribution for the adiabatic compression of a dilute classical gas''
Phys.\ Rev.\ {\bf E 75}, 021116 (2007), arXiv:cond-mat/0603116.
    
  \bi{lua}
  R.C. Lua, A.Y. Grosberg
  ``On practical applicability of the Jarzynski relation in 
   statistical   mechanics: a pedagogical example'', J.\ Phys.\ Chem.
   \ B.,   109(14), 6805 (2005);   arXiv:cond-mat/0502434.

 \bi{relative}
    U. Seifert
   ``Entropy production along a stochastic trajectory and an 
    integral  fluctuation theorem'' 
     Phys.\ Rev.\ Lett.\ {\bf 95}, 040602 (2005);
  
\bibitem{kirone}
For  pedagogic lectures (in French) on work identities and generalizations, see\\
 K.~Mallick,``D�veloppements r�cents en Physique Statistique loin de l'�quilibre''

http://ipht.cea.fr/Docspht/search/article.php?id=t09/325

 \bi{jensen}
J. L. W. V. Jensen,  ``Sur les fonctions convexes et les in�galit�s entre les valeurs moyennes'', Acta Mathematica 30 (1)(1906) 175.

 
\bibitem{standard} 
  A.~H.~Mueller,
  ``The Boltzmann equation for gluons at early times after a heavy ion collision,''
  Phys.\ Lett.\ B {\bf 475}, 220 (2000)
  [hep-ph/9909388].\\
 A.~Krasnitz and R.~Venugopalan,
  ``The Initial gluon multiplicity in heavy ion collisions,''
  Phys.\ Rev.\ Lett.\  {\bf 86}, 1717 (2001)
  [hep-ph/0007108].
  

\bi{avsar1}
  E.~Avsar,
  ``TMD factorization and the gluon distribution in high energy QCD,''
  arXiv:1203.1916 [hep-ph].


\bi{avsar2}
  E.~Avsar and J.~C.~Collins,
  ``Inability to find justification of a $k_T$-factorization formula by following chains of citations,''
  arXiv:1209.1675 [hep-ph].
  
\bi{marquet}
 See, for definition, discussion  and references on the unintegrated gluon distributions, we have been using :\\
  F.~Dominguez, C.~Marquet, B.~W.~Xiao and F.~Yuan,
  ``Universality of Unintegrated Gluon Distributions at small x,''
  Phys.\ Rev.\  D {\bf 83}, 105005 (2011)
  [arXiv:1101.0715 [hep-ph]], \\ and for a comprehensive review:\\
   F.~Dominguez, PHD thesis, ``Unintegrated Gluon Distributions at Small-x'', 2011\\
   http://academiccommons.columbia.edu/catalog/ac:139284
 

\bi {notes}
  The calculations for the piston compression are done 
  using the Crooks  identity \cite{crooksy}:  R.P., unpublished.

\bibitem{lappi}
  T.~Lappi,
  ``Energy density of the glasma,''
  Phys.\ Lett.\  B {\bf 643}, 11 (2006)
  [arXiv:hep-ph/0606207].

\bi{McV}
  L.~D.~McLerran and R.~Venugopalan,
  ``Computing quark and gluon distribution functions for very large nuclei,''
  Phys.\ Rev.\  D {\bf 49}, 2233 (1994)
  [arXiv:hep-ph/9309289];\\
$ibd.$
``Gluon distribution functions for very large nuclei at small transverse
  momentum,''
  Phys.\ Rev.\  D {\bf 49}, 3352 (1994)
  [arXiv:hep-ph/9311205].
 

  
\bibitem{lappi2}
  T.~Lappi,
  ``Gluon spectrum in the glasma from JIMWLK evolution,''
  Phys.\ Lett.\  B {\bf 703}, 325 (2011)
  [arXiv:1105.5511 [hep-ph]].


\bibitem{Kharzeev} 
  D.~Kharzeev and K.~Tuchin,
  ``From color glass condensate to quark gluon plasma through the event horizon,''
  Nucl.\ Phys.\ A {\bf 753}, 316 (2005)
  [hep-ph/0501234].

\bibitem{epelbaum} 
  T.~Epelbaum and F.~Gelis,
  ``Role of quantum fluctuations in a system with strong fields: Spectral properties and Thermalization,''
  Nucl.\ Phys.\ A {\bf 872}, 210 (2011)
  [arXiv:1107.0668 [hep-ph]].
  
 
 


\end{thebibliography}
\end{document}